# A Behavioral Compact Model of 3D NAND Flash Memory

Shubham Sahay, *Student Member, IEEE* and Dmitri Strukov, *Senior Member, IEEE*

*Abstract*— We present a behavioral compact model of 3D NAND flash memory for integrated circuits and system-level applications. This model is easy to implement, computationally efficient, fast, accurate and effectively accounts for the different parasitic capacitance coupling effects applicable to the 3D geometry of the vertical channel Macaroni body charge-trap flash memory. The model parameter extraction methodology is simple and can be extended to reproduce the electrical behavior of different 3D NAND flash memory architectures (with different page size, dimension, or number of stacked layers). We believe that the developed compact model would equip the circuit designers and system architects with an effective tool for design-exploration of 3D NAND flash memory devices for diverse unconventional analog applications.

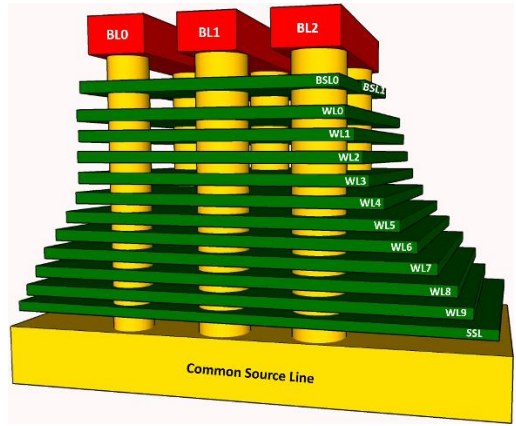

Fig. 1. 3-D view of the vertical charge-trap 3D NAND flash memory array.

## I. INTRODUCTION

NAND flash technology, with the aid of three-dimensional (3D) stacking, continues to be the most promising non-volatile data-storage system to cope up with the overwhelming data explosion in this era of internet-of-things (IoT) [1]-[3]. The 3D NAND flash memory has now become ubiquitous with a wide range of applications ranging from portable USB sticks, camera and smart phone flash drives to solid-state-drives (SSD) [4] and cloud storage [5]. With ultra-high (> 1 Tb) density, ultra-low cost per bit, fast random access and multi-level programming capability per cell [6]-[8], 3D NAND flash also appears lucrative for analog applications requiring dense network of devices, in particular for computing. However, the lack of a proper framework for analyzing the efficacy of 3D NAND flash memory limits the efforts of the designer community in this direction. Although 3D NAND flash string has been widely studied with the help of 3D TCAD simulations [9]-[11], such an approach is slow, computationally expensive and unfeasible for circuit and system-level simulations. Therefore, the development of a simulation program with integrated circuit emphasis (SPICE)-compatible compact model, which could enable the circuit designers and system architects to explore the design-space with 3D NAND flash memory without incurring any computational complexity, is essential. Although several compact models have been proposed for the planar floating gate (FG) NAND flash memory arrays [12]-[16], to the best of our knowledge, no compact model is available for the vertical channel charge-trap (CT) 3D NAND flash memory cells. Moreover, 3D NAND flash memory arrays demand a holistic treatment owing to the significant parasitic coupling attributed to their 3D geometry.

To this end, in this letter, we present a behavioral compact model of the Macaroni body charge-trap 3D NAND flash memory. This model takes into account the different parasitic coupling effects in the memory array owing to the 3D geometry of the cells. Moreover, we formulate a simple two-step model parameter extraction procedure which may be used to reproduce the electrical behavior of different 3D NAND flash memory architectures (with different page size, dimension or number of stacked layers). The developed model is fast, easy to implement, computationally inexpensive and accurately predicts the electrical characteristics of the 3D NAND flash string. We believe that this work is an important step towards exploiting 3D NAND flash for different unconventional analog applications apart from digital memory.

## II. COMPACT MODEL FORMULATION

The three-dimensional view of the 3D NAND flash memory array considered in this work for the formulation of compact model is shown in Fig. 1. It consists of a 3×3 string array of vertical channel charge-trap (CT) devices with a gate stack of oxide/nitride/oxide (O/N/O) and Macaroni body (Fig. 2(a)). Each string consists of 10 metal plate word-lines (WLs) and a select transistor for the bit-line (BSL) and the source-line (SSL), respectively. The basic cell in a string, as shown in Fig.

The authors are with the California Nano Systems Institute (CNSI) and the Department of Electrical and Computer Engineering, University of California, Santa Barbara, California, 93106, U.S.A. This work was supported in part by CRISP, one of six centers in JUMP, a Semiconductor Research Corporation (SRC) program sponsored by DARPA.
(e-mail: shubhamsahay@ucsb.edu)

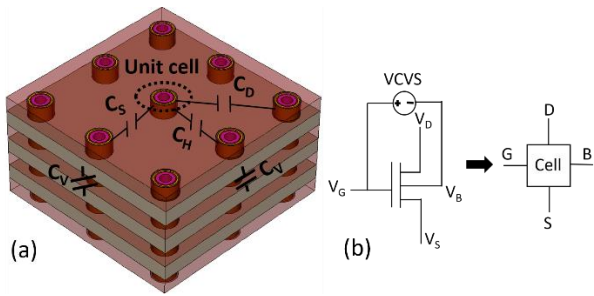

Fig. 2. (a) Different parasitic capacitances in a 3D NAND flash memory array and (b) unit cell representation for compact model development.

TABLE I
PARAMETERS USED FOR 3D NAND BIT STRING SIMULATION

| Parameter | 10 WL string [10] |
|---|---|
| Core filler diameter ($t_f$) | 35 nm |
| Tunnel oxide ($SiO_2$) thickness | 4 nm |
| Blocking oxide ($SiO_2$) thickness | 4 nm |
| Body doping | $1 \times 10^{15}$ cm$^{-3}$ |
| BSL drain doping ($N_{BL}$) | $5 \times 10^{19}$ cm$^{-3}$ |
| SSL source doping ($N_{BL}$) | $5 \times 10^{19}$ cm$^{-3}$ |
| WL work function ($\phi_M$) | 4.8 eV |
| WL length ($L_{WL}$) | 50 nm |
| Spacer ($SiO_2$) thickness ($t_{sp}$) | 50 nm |
| WL thickness ($t_{WL}$) | 40 nm |
| Active channel thickness ($t_{Si}$) | 10 nm |

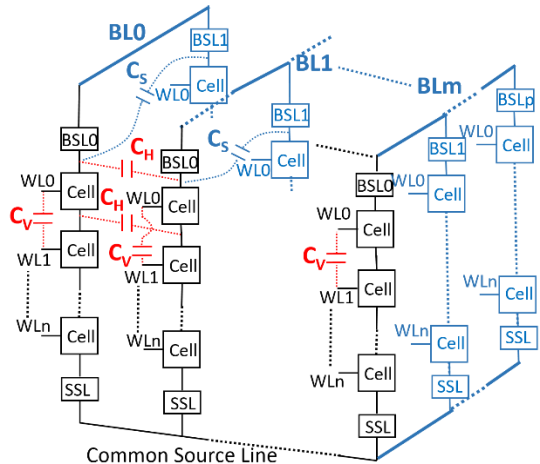

Fig. 3. Representation of a generalized 3D NAND flash memory array utilizing the compact model approach.

2(b), is modelled as a polysilicon gate-all-around nanowire field-effect transistor (GAA-NWFET) with a voltage-controlled-voltage-source (VCVS) to mimic the shift in the threshold voltage upon application of a program/erase pulse. The central string (containing the unit cell in Fig. 2(a)) is analyzed for extraction of the model parameters. Since the parasitic capacitances and the consequent coupling effects are highly pronounced in the 3D NAND flash arrays, the other strings are included to enable accurate estimation of the coupling effects. The cell-cell coupling between adjacent cells sharing the same WL is modelled using a horizontal capacitance ($C_H$) while the vertical coupling between the cells on the same string is taken into account via a vertical capacitance ($C_V$) as shown in Fig. 2(a). In addition, the coupling between diagonal cells and the neighboring cells is considered using the diagonal capacitance ($C_D$) and the side capacitance ($C_S$), respectively. The macro-model representation of a generalized 3D NAND flash array utilizing the compact model approach is shown in Fig. 3. As can be observed from Fig. 3, the 3D NAND flash array demands a comprehensive three-dimensional treatment of different coupling effects for accurate performance estimation as opposed to the planar NAND flash strings [12]-[13].

III. MODEL PARAMETER EXTRACTION METHODOLOGY

We devised a simple parameter extraction methodology for ease of implementation of the compact model without compromising with the accuracy. The extraction procedure is divided into two steps. The first step involves determination of the coupling capacitances of the cell with help of mixed-mode TCAD simulations while the parameters of the basic cell are extracted in the second step.

Mixed-mode simulations of the 3×3×3 string array of vertical channel CT memory devices with O/N/O-stack and Macaroni body (Fig. 2(a)) were carried out using Sentaurus TCAD (release H-2013.03-SP2) [17]. The parameters used for the device simulations are listed in Table I. The different capacitance components ($C_V$, $C_H$, $C_S$ and $C_D$) were extracted using inverted nodal admittance-matrix. The efficacy of the TCAD simulations in yielding an accurate estimation of the geometry dependent electrostatic effects for the flash memory is already established [12], [18]. It may be noted that due to the presence of metal plates for WL, the vertical coupling capacitance ($C_V$) is significantly higher than the other components. Moreover, the diagonal capacitance ($C_D$) is not significant and may be neglected for the cell morphology considered in this work.

Once the coupling capacitances are estimated, the next step is to extract the parameters of the basic cell which consists of a polysilicon GAA-NWFET. The BSIM-CMG 110.0.0 compact model [19] for GAA-NWFET with cylindrical geometry (GEOMOD=3) was used to emulate the cell behavior. BSIM-CMG model parameters may be extracted directly from experimental characterization of the string-current dependence on the WL-voltage of the individual cells located at different positions along the string (single-WL measurement) utilizing the standard procedure reported in [19]. However, due to lack of experimental data for cells at different locations along the string, a two-fold approach was followed in this work. First, the central string (see Fig. 2(a)) with 10 WLs was simulated using Sentaurus TCAD (release H-2013.03-SP2) [17] utilizing the dimensions listed in Table I. The simulated structure (Fig. 4(a)) resembles the Macaroni body vertical channel CT memory obtained using punch-and-plug process utilized for fabricating 3D NAND flash memories [8]-[10]. Although grain boundaries in the polysilicon channel are known to affect the string electrostatics and increase the device variability [9], [11], [20] we have neglected them for the sake of simplicity. Drift-diffusion based simulations were performed assuming a constant mobility and utilizing Shockley-Read-Hall (SRH) recombination model. The TCAD simulation set up was calibrated by reproducing the experimental string current of a similar test structure [9] as shown in Fig. 4(b). The voltages of all 10 WLs were ramped together (multi-WL measurement) to explore the average behavior of the string current [9]. It may be noted that the main aim of this work is to propose a method for formulating a compact model of 3D NAND flash rather than showing the exact values of the string current.

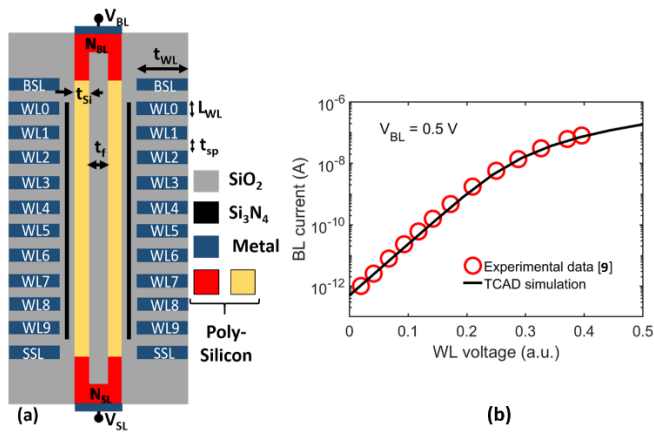

Fig. 4. (a) Cross-sectional view of the central string used for the 3D TCAD simulations and (b) calibration of the TCAD simulation set up by reproducing the experimental string-current multi-WL characteristics of 10 WL string [9].

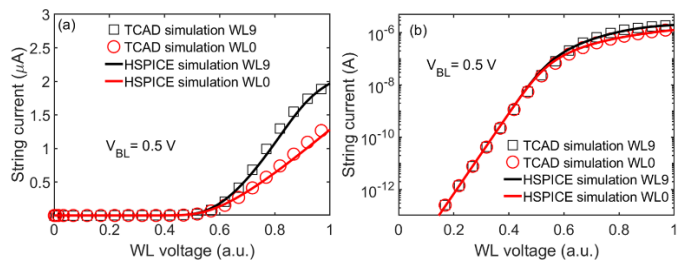

Fig. 5. String current single-WL characteristics of WL0 and WL9 in (a) linear scale and (b) log scale obtained using TCAD simulations and HSPICE simulations after tuning the model parameters of BSIM-CMG 110.0 model.

Subsequently, the calibrated TCAD set up was used to obtain the single-WL measurements for cells located at different positions along the string. The source/drain and channel series resistance is significantly higher in the 3D NAND flash string owing to the lightly doped polysilicon film. Therefore, when probing a particular cell, the WLs of all the other cells were biased at a high voltage ($V_{WL} = 5$ V) to reduce the impact of series resistance. The parameters of the BSIM-CMG model were tuned following the standard extraction procedure [19] to reproduce the TCAD simulation results. As can be observed from Fig. 5, the HSPICE (version N-2017.12 [21]) simulation results of the BSIM-CMG model with modified parameters are in good agreement with the single-WL characteristics obtained via TCAD simulations for cells located at the extreme ends of the string: WL0 at the top and WL9 at the bottom. This clearly indicates the efficacy of the proposed model parameter extraction procedure in reproducing the electrical behavior of the 3D NAND flash string. The program/erase capabilities can also be appended to the compact model by utilizing a voltage-controlled-voltage source (VCVS) as shown in Fig. 2(a) and calibrating the parameters of the VCVS to match the experimental measurements.

It may be noted that we have considered a string with only 10 WLs in this work owing to the lack of published experimental data for strings with higher number of stacked WLs. However, the proposed approach may be extended to model 3D NAND flash memory arrays with different page size, dimension, or number of stacked WL layers.

## IV. CONCLUSIONS

In this letter, we have formulated a behavioral compact model for the 3D NAND flash memory. The simple model parameter extraction procedure is described in detail. The model accurately reproduces the electrical behavior of the string taking into account the capacitive coupling in different directions inherent to the 3D geometry. This work is an important step in the direction of exploiting 3D NAND flash memories for different unconventional analog applications.

REFERENCES

[1] H. Kim, S. J. Ahn, Y. G. Shin, K. Lee, and E. Jung, "Evolution of NAND Flash Memory: from 2D to 3D as a storage market leader," in *IEEE Int. Mem. Work.*, pp. 1-4, May. 2017. doi: 10.1109/IMW.2017.7939081.
[2] C. Monzio Compagnoni, A. Goda, A. S. Spinelli, P. Feeley, A. L. Lacaita, and A. Visconti, "Reviewing the evolution of the NAND Flash technology," *Proc. IEEE*, vol. 105, no. 9, pp. 1609–1633, Sep. 2017, doi: 10.1109/JPROC.2017.2665781.
[3] R. Micheloni, S. Aritome, and L. Crippa, "Array architectures for 3-D NAND flash memories," *Proc. IEEE*, vol. 105, no. 9, pp. 1634-1649, Sep. 2017, doi: 10.1109/JPROC.2017.2697000.
[4] L. Zuolo, C. Zambelli, R. Micheloni, and P. Olivo, "Solid-state drives: memory driven design methodologies for optimal performance," *Proc. IEEE*, vol. 105, no. 9, pp.1589-1608, Sep. 2017, doi: 10.1109/JPROC.2017.2733621.
[5] D. Narayanan, E. Thereska, A. Donnelly, S. Elnikety, and A. Rowstron, "Migrating server storage to SSDs: analysis of tradeoffs," in *Proc. ACM Eur. Conf. Comp. syst.*, pp. 145-158, Apr. 2009, doi: 10.1145/1519065.1519081.
[6] T. Tanaka, M. Helm, T. Vali, R. Ghodsi, K. Kawai, J. K. Park, S. Yamada, F. Pan, Y. Einaga, A. Ghalam, and T. Tanzawa, "A 768 Gb 3b/cell 3D-floating-gate NAND Flash memory," in *ISSCC Dig. Tech. Papers*, Jan. 2016, pp. 142–143, doi: 10.1109/ISSCC.2016.7417947.
[7] S. Lee, C. Kim, M. Kim, S. M. Joe, J. Jang, S. Kim, K. Lee, J. Kim, J. Park, H. J. Lee, and M. Kim, "A 1Tb 4b/cell 64-stacked-WL 3D NAND flash memory with 12MB/s program throughput,". in *ISSCC Dig. Tech. Papers*, pp. 340-342, Feb. 2018, doi: 10.1109/ISSCC.2018.8310323.
[8] H. Tanaka, M. Kido, K. Yahashi, M. Oomura, R. Katsumata, M. Kito, Y. Fukuzumi, M. Sato, Y. Nagata, Y. Matsuoka, and Y. Iwata, "Bit cost scalable technology with punch and plug process for ultra high density flash memory," in *VLSI Symp. Tech. Dig.*, pp. 1–2, 2007, doi: 10.1109/VLSIT.2007.4339708.
[9] D. Resnati, A. Mannara, G. Nicosia, G. M. Paolucci, P. Tessariol, A. S. Spinelli, A. L. Lacaita, and C. M. Compagnoni, "Characterization and modeling of temperature effects in 3-D NAND Flash arrays—Part I: Polysilicon-induced variability," *IEEE Trans. Electron Devices*, vol. 65, no. 8, pp.3199-3206, 2018, doi: 10.1109/TED.2018.2838524.
[10] G. Malavena, A. L. Lacaita, A. S. Spinelli, and C. M. Compagnoni, "Investigation and compact modeling of the time dynamics of the GIDL-assisted increase of the string potential in 3-D NAND flash arrays," *IEEE Trans. Electron Devices*, vol. 65, no. 7, pp.2804-2811, Jul. 2018, doi: 10.1109/TED.2018.2831902.
[11] D. Resnati, A. Mannara, G. Nicosia, G. M. Paolucci, P. Tessariol, A. L. Lacaita, A. S. Spinelli, and C. M. Compagnoni, "Temperature activation of the string current and its variability in 3-D NAND Flash arrays," in *IEDM Tech. Dig.*, Jan. 2017, pp. 4.7.1–4.7.4, doi: 10.1109/IEDM.2017.8268329.
[12] L. Larcher, A. Padovani, P. Pavan, P. Fantini, A. Calderoni, A. Mauri, and A. Benvenuti, "Modeling NAND Flash memories for IC design," *IEEE Electron Device Lett.*, vol. 29, no. 10, pp.1152-1154, Oct.2008, doi: 10.1109/LED.2008.2003179.
[13] A. Spessot, C. M. Compagnoni, F. Farina, A. Calderoni, A. S. Spinelli, and P. Fantini, "Compact modeling of variability effects in nanoscale

NAND flash memories," *IEEE Trans. Electron Devices*, vol. 58, no. 8, pp.2302-2309, Aug 2011, doi: 10.1109/TED.2011.2147319.

[14] J. Jeon, I. H. Park, M. Kang, W. Hahn, K. Choi, S. Yun, G. Y. Yang, K. H. Lee, Y. K. Park, and C. Chung, "Accurate compact modeling for sub-20-nm NAND flash cell array simulation using the PSP model," *IEEE Trans. Electron Devices*, vol. 59, no. 12, pp.3503-3509, Dec. 2012, doi: 10.1109/TED.2012.2219863

[15] A. Maure, P. Canet, F. Lalande, B. Delsuc, and J. Devin, "Flash memory cell compact modeling using PSP model," in *IEEE Int. Beh. Mod. Sim. Work. (BMAS)*, pp. 45-49, Sep 2008, doi: 10.1109/BMAS.2008.4751238

[16] M. Kang, W. Hahn, I. H. Park, Y. Song, H. Lee, K. Choi, Y. Lim, S. M. Joe, D. H. Chae, and H. Shin, "A compact model for channel coupling in sub-30 nm NAND flash memory device," *Jap. J. App. Phy.*, vol. 50, no. 10R, p.100204, Oct 2011, doi: 10.1143/JJAP.50.100204.

[17] *Sentaurus Device User Guide*, Synopsys, Inc., Mountain View, CA, 2014.

[18] A. Ghetti, L. Bortesi, and L. Vendrame, "3D simulation study of gate coupling and gate cross-interference in advanced floating gate nonvolatile memories," *Solid-State Electron.*, vol. 49, no. 11, pp. 1805–1812, Nov. 2005, doi: 10.1016/j.sse.2005.10.014.

[19] *BSIM-CMG 110.0.0 Technical Manual*., Jan, 2016, [online] Available: www.bsim.berkeley.edu/models/bsimcmg/

[20] A. Subirats, A. Arreghini, E. Capogreco, R. Delhougne, C. L. Tan, A. Hikavyy, L. Breuil, R. Degraeve, V. Putcha, D. Linten, and A. Furnémont, "Experimental and theoretical verification of channel conductivity degradation due to grain boundaries and defects in 3D NAND," in *proc. IEEE Int. Elec. Dev. Meet. (IEDM)*, pp. 21-22, Dec. 2017, doi: 10.1109/IEDM.2017.8268433.

[21] *HSPICE user guide: basic simulation and analysis*, Synopsys, Inc., Mountain View, CA, 2018.